\documentclass{article}
\usepackage{graphicx}
\usepackage{epsfig}
\usepackage{amsmath}
\usepackage{authblk}
\usepackage{caption}
\usepackage{cite}
\usepackage{color}
\usepackage{epstopdf}
\usepackage{float}
\usepackage{geometry}
\usepackage{graphicx}
\usepackage{hyperref}
\usepackage{listings}
\usepackage{textcomp}
\lstdefinelanguage{partialsagemathmaxima}
{
  morekeywords=
  {
  reset, from, import, var, assume, print, show, inverse, display, expr, function, for, range, len, diff, numerator, exp, simplify_full,kill,if,get,then,load,ratwtlvl,ratfac,matrix,invert,else,in,expand,collect,save,subs,
  },
  sensitive=false,
  morecomment=[l]{\#},
}
\begin{document}
%%%%%%%%%%%%%%%%%%%%%% 
\title{Symbolic and Numerical Analysis in General Relativity with Open Source Computer Algebra Systems}
\author[$1$]{Tolga Birkandan \footnote{E-mail address: birkandant@itu.edu.tr}}
\author[$1$]{Ceren G\"{u}zelg\"{u}n} 
\author[$1$]{Elif \c{S}irin} 
\author[$2$]{Mustafa Can Uslu}
\affil[$1$]{Istanbul Technical University, Department of Physics, Istanbul, Turkey}
\affil[$2$]{Istanbul Technical University, Institute of Informatics, Istanbul, Turkey}
%%%%%%%%%%%%%%%%%%%%%%%
%\title{Symbolic and Numerical Analysis in General Relativity with Open Source Computer Algebra Systems}
%\author{Tolga Birkandan \and Ceren G\"{u}zelg\"{u}n \and Elif \c{S}irin \and Mustafa Can Uslu}
%
%\institute{Tolga Birkandan \and Ceren G\"{u}zelg\"{u}n \and Elif \c{S}irin  \at
%	Istanbul Technical University, Department of Physics, Istanbul, Turkey \\
%	\email{birkandant@itu.edu.tr}           %  \\
%	\and
%	Mustafa Can Uslu \at
%	Istanbul Technical University, Institute of Informatics, Istanbul, Turkey
%}
%
%\date{Received: date / Accepted: date}

%%%%%%%%%%%%%%%%%%%%%%%
%\begin{document}
\maketitle
%%%%%%%%%%%%%%%%%%%%%%%
\begin{abstract}
\noindent We study three computer algebra systems, namely SageMath (with SageManifolds package), Maxima (with ctensor package) and Python language (with GraviPy module), which allow tensor manipulation for general relativity calculations along with general algebraic calculations. We present a benchmark of these systems using simple examples. After the general analysis, we focus on the SageMath and SageManifolds system to derive, analyze and visualize the solutions of the massless Klein-Gordon equation and geodesic motion with Hamilton-Jacobi formalism. We compare our numerical result of the Klein-Gordon equation with the asymptotic form of the analytical solution to see that they agree.
\end{abstract}
%\newpage
%\tableofcontents
%\newpage
%%%%%%%%%%%%%%%%%%%%%% 
\section{Introduction}
Computer algebra systems are essential tools for theoretical physics for some decades. They are mainly important in general relativity (GR) where lengthy tensorial and differential geometry calculations are inevitable. Many of these programs have special internal or external packages for tensor manipulation and differential geometry calculations. For comprehensive reviews, see \cite{Heinicke:2001ug,Korolkova:2014oia,MacCallum:2018csx}. The review of early works can be found in \cite{aman}. These packages can also be extended for more specialized calculations \cite{Birkandan:2007cx}.

Even though commercial programs dominate the area, codes written on commercial programs cannot be distributed easily since those programs may not be available for scientists with a lack of resources for purchasing such software. Besides, many of the manipulations treated in calculations do not need a sophisticated computation engine.

Open source computer algebra systems such as SageMath (also known as ``Sage") \cite{sage} and Maxima \cite{maxima} provide a complete toolkit for general relativity and quantum field theory applications with their particular packages. Some freely available programming languages such as Python \cite{python} also offer special tools for the aforementioned manipulations.

We choose to study general systems rather than specialized tensor manipulation or general relativity packages (e.g. Cadabra \cite{cadabra}, Redberry \cite{redberry}, etc.). The reason for this choice is that tensor manipulation is generally an auxiliary step in GR calculations. A general program which can deal with the symbolic and numerical analysis of the results and produce graphical outputs constitute a complete calculation toolkit. In most cases, computer algebra systems (or programming languages) are supported with particular packages for tensor manipulation. We also exclude specialized tools like GYOTO \cite{gyoto} in our analysis.

In this work, we will first employ SageMath (with SageManifolds package \cite{Gourgoulhon:2014ywa}), Maxima (with ctensor package \cite{toth}), and Python language (with Sympy and GraviPy modules \cite{gravipy}) for some essential calculations in general relativity and present benchmark results for these systems.

The ever-developing open source SageMath program has gathered many utilities such as Maxima, GAP, R, and the power of Python language with well-known Python modules like NumPy, SymPy and matplotlib. SageMath can be installed on personal computers and moreover it has a powerful cloud computing server on which the user can work on projects anywhere and share them with other users easily \cite{sage}. The package SageManifolds for tensor and differential geometry calculations is included in SageMath and it does not require an additional installation process. These properties make the Sage+SageManifolds system one of the best open source choices for general relativity and quantum field theory. A comprehensive lecture note on SageManifolds is available in \cite{Gourgoulhon:2018yss}.

Analysis of the Klein--Gordon equation, mainly its radial part is the first step in most quantum gravity problems involving black holes \cite{Birkandan:2011fr}. We will study its solution using SageMath and see that the numerical result found by SageMath agrees with the asymptotic analytical result.

Computationally, it is easier to define and solve the first order differential equations rather than the second order equations. Hamilton--Jacobi formalism yields first order differential equations for the geodesic motion \cite{chandr}. We will use this formalism to derive the equations and solve them as a numerical initial value problem with SageMath routines. Numerical results need to be presented in a graphical way to see their structure efficiently and SageMath system is equipped with many visualization tools. The ``Examples" section in SageManifolds web page \cite{Gourgoulhon:2014ywa} also presents some applications on geodesics. However, our code provides a step-by-step procedure for the Hamilton-Jacobi approach to geodesic motion with explicit numerical analysis.

In the next section, we define the spacetimes to be used. In the third section we review three computation systems and give simple examples. We also give a benchmark of these systems as a subsection. In the fourth section we focus on the SageMath+SageManifolds system to analyze the massless Klein--Gordon equation and geodesic motion.

All codes studied in this paper can be accessed from the GitHub address \cite{github}, organized by branches.
%%%%%%%%%%%%%%%%%%%%%%
\section{Schwarzschild and Kerr solutions}
We will be working for Schwarzschild and Kerr spacetimes with the metric signature $(+,-,-,-)$. In general, a line element is defined by
\begin{equation} 
ds^2= g_{\mu\nu}dx^{\mu}dx^{\nu}.
\end{equation}
If one takes $G_4=c=1$ (where $G_4$ is Newton's gravitational constant in four dimensions and $c$ is the speed of light in vacuum), the Schwarzschild metric can be written as \cite{chandr}
\begin{equation} 
g_{\mu\nu}=\left[\begin{array}{cccc}
\Big( 1- \frac{2M}{r} \Big) & 0 & 0 & 0	\\
0 & -\Big( 1- \frac{2M}{r} \Big)^{-1} & 0 & 0	\\
0 & 0 & -r^2 & 0	\\
0 & 0 & 0 & -r^2sin^2\theta
\end{array}\right]. \label{schwmetric}
\end{equation}
We will take the order of coordinates as $\{t,r,\theta,\phi\}$ and $M$ is the mass of the black hole. The coefficient of the radial part is singular at $r=2M$ which describes the ``event horizon". 

In the Boyer-Lindquist coordinates, the Kerr black hole has the metric \cite{chandr}
\begin{equation} 
g_{\mu\nu}=\left[\begin{array}{cccc}
\Big( 1- \frac{2M}{\rho^2} \Big) & 0 & 0 & \frac{2aMrsin^2\theta}{\rho^2}	\\
0 & -\frac{\rho^2}{\Delta} & 0 & 0	\\
0 & 0 & -\rho^2 & 0	\\
\frac{2aMrsin^2\theta}{\rho^2} & 0 & 0 & -sin^2\theta \big[(r^2+a^2)+\frac{2a^2Mrsin^2\theta}{\rho^2} \big]
\end{array}\right]. \label{kerrmetric}
\end{equation}
Here, $M$ is the mass and $a=J/M$ where $J$ is the angular momentum. Two functions are defined as
\begin{align} 
\rho^2&=r^2+a^2cos^2\theta, \\
\Delta&=r^2-2Mr+a^2.
\end{align}
The roots of $\Delta=0$ correspond to the locations of the Cauchy horizon (the smaller root) and the event horizon (the larger root) of the black hole.
%%%%%%%%%%%%%%%%%%%%%%
\section{Open source computer algebra systems for basic GR calculations}
We will only focus on the procedures of
\begin{itemize}
	\item Defining a spacetime
	\item Calculating Christoffel symbols ($\Gamma^{\mu}_{\nu \rho}$)
	\item Calculating Ricci tensor ($R_{\mu \nu}$) 
	\item Calculating Einstein tensor ($G_{\mu \nu}$)
	\item Displaying components 
\end{itemize}
in our examples. The examples will be studied using three computation systems, namely
\begin{itemize}
	\item SageMath+SageManifolds
	\item Maxima+ctensor
	\item Python+GraviPy
\end{itemize}
The codes are enhanced with comment lines which can be helpful for analyzing the procedures step by step. Details on the procedures and further examples can be found in related references.
%%%%%%%%%%%%%%%%%%%%%%
\subsection{SageMath+SageManifolds}
SageMath is an open source computer algebra system which collects many powerful open source packages and modules with a Python-like language \cite{sage}. SageManifolds was first started as an independent tensor analysis and differential geometry package to be installed on SageMath \cite{Gourgoulhon:2014ywa}. Now, SageManifolds is an internal package for SageMath and does not require an additional installation.

We start our program by reseting the SageMath environment. Then we define the four-dimensional manifold, the black hole mass and the coordinates. The coordinates are defined along with their ranges. We define the Lorentzian metric by giving its components as a matrix. Finally, we calculate and display Christoffel symbols, Ricci tensor and Einstein tensor elements.
\begin{lstlisting}[frame=single]
reset()
# Define 4-dim. the manifold "Man":
Man = Manifold(4, 'Man', r'\mathcal{M}')

# Define the parameter "M" (mass):
M = var('M')

# Define the coordinates {t=0, r=1, theta(=th)=2, phi=3} with ranges
# (BL = Boyer-Lidquist)
BL.<t,r,th,ph> = Man.chart(r't r:(0,+oo) th:(0,pi):\theta ph:(0,2*pi):\phi')

# Define the metric "g" on manifold "Man":
g = Man.lorentzian_metric('g')

# Enter the metric components:
g[0,0] = (1-(2*M)/r) (*@\label{schwstart}@*) 
g[1,1] = -1/(1-(2*M)/r)
g[2,2] = -r^2
g[3,3] = -(r*sin(th))^2 (*@\label{schwstop}@*)

# Display the metric
show('The metric:')
show(g.display())
#####################
# Christoffel symbols
nab = g.connection()
# Display all components
show(nab.display())
# Display a single component
show(nab[3,2,3])
#####################
# Ricci tensor
Ric = g.ricci()
# Display all components
show(Ric.display())
# Display a single component
show(Ric[1,1])
#####################
# Einstein tensor
ET=g.ricci()-(1/2)*g*g.ricci_scalar()
# Display all components
ET.set_name('E')
show(ET.display())
# Display a single component
show(ET[1,2])
\end{lstlisting}
The lines between \ref{schwstart}-\ref{schwstop} defines the Schwarzschild metric. In order to define the Kerr metric, we need to change this part with
\begin{lstlisting}[frame=single,firstnumber=16]
a = var('a')
rhosq=r^2+(a^2)*cos(th)^2
Delta=r^2-2*M*r+a^2

g[0,0] = (1-(2*M*r)/rhosq)
g[1,1] = -rhosq/Delta
g[2,2] = -rhosq
g[3,3] = -(sin(th)^2)*((r^2+a^2)+(2*(a^2)*M*r*sin(th)^2)/rhosq)
g[0,3] = (2*a*M*r*sin(th)^2)/rhosq
\end{lstlisting}
It should be noted that, we defined the functions $\rho^2$ and $\Delta$, and the rotation parameter $a$. The Kerr black hole has non-diagonal, symmetric components unlike the Schwarzschild metric which is diagonal.
%%%%%%%%%%%%%%%%%%%%%%
\subsection{Maxima+ctensor}
Maxima computer algebra system is a freely available descendant of Macsyma which is known as the first general, multipurpose computer algebra system which inspired many other systems for years \cite{maxima}. V. Toth reviews tensor packages in Maxima in his article \cite{toth}. Here, we will use the ctensor package which provides component tensor manipulation.

We first kill all environmental variables, then load the ctensor package. We set {\tt ratwtlvl:false} for no truncation and {\tt ratfac:true} to factorize the tensor components automatically. We then define the dimension of the spacetime and the coordinates. We enter the metric as a matrix. Before the calculation of Christoffel symbols, Ricci tensor and Einstein tensor elements, we find the inverse tensor with {\tt invert}.
\begin{lstlisting}[frame=single]
kill(all);
if get('ctensor,'version)=false then load(ctensor);
(ratwtlvl:false,ratfac:true);
("Specify the dimension of the manifold and the coordinate labels.")$
(dim:4,ct_coords:[t,r,theta,phi]);
("Enter the metric.")$ (*@\label{schwmax1}@*)
lg:matrix([(1-2*M/r),0,0,0],[0,-1/(1-2*M/r),0,0],[0,0,-r^2,0],[0,0,0,-(r^2)*sin(theta)^2]); (*@\label{schwmax2}@*)
ug:invert(lg)$
("Compute the Christoffel symbols and display all components")$
christof(mcs)$
("Compute the Ricci tensor and display all components")$
uricci(true)
("Compute the Einstein tensor and display all components")$
einstein(true);
("Display single components")$
mcs[3,4,4];
ric[1,2];
ein[2,2];
\end{lstlisting}
The lines \ref{schwmax1}-\ref{schwmax2} should be changed as
\begin{lstlisting}[frame=single,firstnumber=6]
rhosq:r^2+(a^2)*cos(theta)^2$
Delta:r^2-2*M*r+a^2$
("Enter the general static spherically symmetric metric.")$
lg:matrix([(1-(2*M*r)/rhosq),0,0,(2*a*M*r*sin(theta)^2)/rhosq],[0,-rhosq/Delta,0,0],[0,0,-rhosq,0],[(2*a*M*r*sin(theta)^2)/rhosq,0,0,-(sin(theta)^2)*((r^2+a^2)+(2*(a^2)*M*r*sin(theta)^2)/rhosq)]);
\end{lstlisting}
for the Kerr metric.
%%%%%%%%%%%%%%%%%%%%%%
\subsection{Python+GraviPy}
Python is a multipurpose, object-oriented programming language \cite{python} which can easily be expanded with modules. The module GraviPy provides tensor calculation methods and it works on a freely available symbolic analysis module SymPy \cite{gravipy}.

We start our program by importing the GraviPy module. Then we define the four--vector of coordinates and the black hole mass. We define the metric as a diagonal matrix. We then calculate and display Christoffel symbols, Ricci tensor and Einstein tensor elements.
\begin{lstlisting}[frame=single]
from gravipy import *
##################### (*@\label{schwpy1}@*)
# Coordinates (\chi is the four-vector of coordinates)
t, r, theta, phi, M = symbols('t, r, theta, phi, M') 
x = Coordinates('\chi', [t, r, theta, phi])
#####################
# Metric tensor
Metric = diag((1-2*M/r), -1/(1-2*M/r), -r**2, -r**2*sin(theta)**2) (*@\label{schwpy2}@*)
g = MetricTensor('g', x, Metric)
#####################
# Christoffel symbols
Ga = Christoffel('Ga', g)
# Display all components
print(Ga(All, All, All))
# Display a single component
print(Ga(1,2,1))
#####################
# Ricci tensor
Ri = Ricci('Ri', g)
# Display all components
print(Ri(All, All))
# Display a single component
print(Ri(1, 2))
#####################
# Einstein tensor
G = Einstein('G', Ri)
# Display all components
print(G(All, All))
# Display a single component
print(G(3, 3))
#####################
\end{lstlisting}
The lines \ref{schwpy1}-\ref{schwpy2} should be modified as
\begin{lstlisting}[frame=single,firstnumber=2]
#####################
# Coordinates (\chi is the four-vector of coordinates)
t, r, theta, phi, M, a, rhosq, Delta = symbols('t, r, theta, phi, M, a, rhosq, Delta') 
x = Coordinates('\chi', [t, r, theta, phi])
#####################
# Metric tensor
rhosq = r**2+(a**2)*cos(theta)**2
Delta = r**2-2*M*r+a**2
Metric = Matrix([[(1-(2*M*r)/rhosq),0,0,(2*a*M*r*sin(theta)**2)/rhosq],[0,-rhosq/Delta,0,0],[0,0,-rhosq,0],[(2*a*M*r*sin(theta)**2)/rhosq,0,0,-(sin(theta)**2)*((r**2+a**2)+(2*(a**2)*M*r*sin(theta)**2)/rhosq)]])
\end{lstlisting}
in order to define the Kerr metric. The diagonal matrix definition of the Schwarzschild case is changed with a general metric.
%%%%%%%%%%%%%%%%%%%%%%
\subsection{Benchmark for the open source computer algebra systems}
We performed some calculations for the Schwarzschild and Kerr spacetimes on Python+GraviPy (Python 2.7.15), SageMath+SageManifolds (SageMath 8.3) and Maxima+ctensor (Maxima 18.02.0) systems. 

The Schwarzschild metric has only diagonal elements, while Kerr solution has non--diagonal elements in its metric. We aimed to measure the effect of this difference in the computations. We used the metric given in equation \ref{schwmetric} for the Schwarzschild case and for the Kerr metric the metric is given in equation \ref{kerrmetric}.

For our analysis, we used the code--block profiling method and measured the wall--clock time as it has more importance for the general user. We calculated the Christoffel symbols, Ricci tensor and Einstein tensor one by one on each system. Metric definition and displaying the results are not included in the time measurement. By averaging the timing results for 10 runs on each system, the Table \ref{bench} is generated.

In Python, we used the {\tt datetime} module. This method shows the time value up to six decimals.
\begin{lstlisting}[frame=single]
from gravipy import *
from datetime import datetime
...
riccitime_=datetime.now()
Ri=Ricci('Ri',g)
_riccitime=datetime.now()
print "Elapsed time...", _riccitime-riccitime_
...
\end{lstlisting}
In SageMath, we employed the inline {\tt timeit} command.
\begin{lstlisting}[frame=single]
...
timeit(g.ricci())
...
\end{lstlisting}
In Maxima, we used {\tt elapsed\_real\_time()} command which also shows time with two decimals.
\begin{lstlisting}[frame=single]
...
t0:elapsed_real_time()$
uricci(false);
t1:elapsed_real_time()-t0;
...
\end{lstlisting}
The test computer has an Intel(R) Core(TM) i7-4930K CPU @ 3.40GHz, 16 GB (DIMM DDR3 Synchronous 1066 MHz) RAM, NVIDIA GeForce GTX 650 Ti Boost GPU, 120GB SSD and 2 TB HDD with Ubuntu 18.04.
\begin{table}[H]
	\centering
	\begin{tabular}{|l|l|l|l|l|}
		\hline
		Metric - Calculation & Python & \vtop{\hbox{\strut SageMath}}  &  Maxima   \\ 
		\hline
		Schwarzschild - Christoffel  & 0.000047 & 0.0000030   & 0.10 \\ 
		\hline
		Schwarzschild - Ricci        & 0.000026 & 0.0000024   & 0.09 \\ 
		\hline
		Schwarzschild - Einstein     & 0.000076 & 0.174    & 0.09 \\
		\hline
		Kerr - Christoffel           & 0.000049 & 0.0000028     & 0.20 \\ 
		\hline
		Kerr - Ricci                 & 0.000028 & 0.0000024   & 1.67 \\ 
		\hline
		Kerr - Einstein              & 0.000080 & 0.286  & 2.53 \\ 
		\hline
	\end{tabular}
		\caption{Benchmark results. (All numbers are in seconds).}
		\label{bench}
\end{table}
According to the benchmark results, Python seems to be the best choice for these calculations. However, the calculation commands lack the simplification routines that are needed for a useful result. For example, displaying even one component of the Einstein tensor of the Kerr black hole is impossible in an acceptable duration in Python. 

Focusing on the Kerr solution, we see that SageMath would be the best option with its speed and yielding convenient results for further calculations and manageable outputs.
%%%%%%%%%%%%%%%%%%%%%%
\section{Scalar wave equation and geodesics with SageMath and SageManifolds}
SageMath, being powered by SageManifolds, provides an easy-to-use and combined toolkit for the general user. We will focus on the SageMath+SageManifolds system to perform a set of calculations for the Schwarzschild metric. 

We will first define the spacetime by declaring its variables and components. Then we will perform calculations for two examples. In the first one, we derive the Klein--Gordon equation for a massless scalar field, extract its radial part and solve this differential equation numerically. We will compare our numerical result graphically with the asymptotic form of the analytical result. In the next example, we will perform a very simple geodesic analysis for this spacetime using the Hamilton--Jacobi formalism.

Generalization of the example codes to other metrics is straightforward. The codes for the calculation of Klein--Gordon and Hamilton--Jacobi equations can be used for general metrics without modification. For the detailed calculations, we applied some metric-related information in the code to see the manipulation, equation solving and plotting capabilities of the computation system.

The reader should follow the line numbers to execute the codes without problem. The code for the definition of the spacetime is between lines (\ref{ex1}-\ref{ex2}). The analysis of the Klein--Gordon equation starts in line \ref{ex3} and ends in line \ref{ex4}. Then in the next section, the study of geodesics starts again with line \ref{ex5} and ends in line \ref{ex6}. This means that, the reader should first execute the lines (\ref{ex1}-\ref{ex2}) for the metric definition before each physical example.
%%%%%%%%%%%%%%%%%%%%%%
\subsection{Definition of the spacetime in SageManifolds}
This part was studied before in Section 3.1, while giving examples for simple calculations in GR. Nevertheless, we will place it here in order to present a complete code structure. 
\begin{lstlisting}[frame=single][name=metricdef]
reset() (*@\label{ex1}@*)

# Define 4-dim. the manifold "Man":
Man = Manifold(4, 'Man', r'\mathcal{M}')

# Define the parameter "M" (mass):
M = var('M')

# Define the coordinates {t=0, r=1, theta(=th)=2, phi=3} with ranges
# (BL = Boyer-Lidquist)
BL.<t,r,th,ph> = Man.chart(r't r:(0,+oo) th:(0,pi):\theta ph:(0,2*pi):\phi')

# Define the metric "g" on manifold "Man":
g = Man.lorentzian_metric('g')

# Enter the Schwarzschild metric components:
g[0,0] = (1-(2*M)/r)
g[1,1] = -1/(1-(2*M)/r)
g[2,2] = -r^2
g[3,3] = -(r*sin(th))^2

# Display the metric
show('The Schwarzschild metric:')
show(g.display()) (*@\label{ex2}@*)
\end{lstlisting}
In the Schwarzschild case, the manifold has four dimensions and the only variable is the black hole mass $M$. We define four spacetime coordinates and enter the components of the diagonal metric.
%%%%%%%%%%%%%%%%%%%%%%
\subsection{Klein--Gordon equation in curved spacetime}
Klein--Gordon equation for a massless scalar field can be written as \cite{birrell}
\begin{equation} \label{520}
\frac{1}{\sqrt{-g}}\partial_\mu (\sqrt{-g}g^{\mu\nu} \partial_\nu \Phi) = 0,
\end{equation}
where $\Phi$ is the scalar field and it can be decomposed with the Ansatz
\begin{equation}
\Phi(t,r,\theta,\phi) = e^{-i \omega t} e^{ik\phi} \: R(r) \: S(\theta),
\end{equation}
for the Schwarzschild metric.

We will first define the variables $\omega$, $k$ and the result {\tt KG}. The inverse metric and $\sqrt{-g}$ are present in the equation, thus are calculated. Then the full scalar function $\Phi$, radial and angular parts of the solution Ansatz are defined. After giving the Ansatz, we start calculating the Klein--Gordon equation in two {\tt for} loops. The outer loop is over $\mu$ components and the inner one is over $\nu$ components.
\begin{lstlisting}[frame=single,firstnumber=last]
# Defining variables: (*@\label{ex3}@*)
var('omega,k,KG')

# Inverse metric:
ginv = g.inverse()

# The square root of the absolute value 
# of the metric determinant:
sqrtabsdetg=g.sqrt_abs_det().expr()

# The scalar function Phi(t,r,th,phi):
# The dependence on all coordinates 
# is provided by "(*BL)" 
Phi=function('Phi')(*BL)

# The scalar field Ansatz is given here.
# R : Radial part,
# S : Angular part
R=function('R')(r)
S=function('S')(th)
Phi=exp(-I*omega*t)*exp(I*k*ph)*R*S

# Calculating the Klein-Gordon equation:
KG=0
for mu in range(len(BL[:])):
	for nu in range(len(BL[:])):
		KG=KG+diff((ginv[mu,nu].expr()*sqrtabsdetg*diff(Phi,BL[nu])),BL[mu])

# Displaying the Klein-Gordon equation "KG"
show('The full Klein-Gordon equation (variable name is KG):')
show(KG)
\end{lstlisting}
This part of the code is general. The reader should only change the solution Ansatz accordingly and the code can find the Klein-Gordon equation for any spacetime with any number of dimensions. The result is stored in variable {\tt KG}.

In our example, the Klein--Gordon equation is found as
\begin{align}
\texttt{KG}=&\frac{\omega^{2} r^{3} R\left(r\right) S\left({\theta}\right) e^{\left(i \, k {\phi} - i \, \omega t\right)} \sin\left({\theta}\right)}{2 \, M - r} + {\left(2 \, M - r\right)} r S\left({\theta}\right) e^{\left(i \, k {\phi} - i \, \omega t\right)} \sin\left({\theta}\right) \frac{\partial^{2}}{(\partial r)^{2}}R\left(r\right) \nonumber\\
+& {\left(2 \, M - r\right)} S\left({\theta}\right) e^{\left(i \, k {\phi} - i \, \omega t\right)} \sin\left({\theta}\right) \frac{\partial}{\partial r}R\left(r\right) - r S\left({\theta}\right) e^{\left(i \, k {\phi} - i \, \omega t\right)} \sin\left({\theta}\right) \frac{\partial}{\partial r}R\left(r\right) \nonumber\\
+& \frac{k^{2} R\left(r\right) S\left({\theta}\right) e^{\left(i \, k {\phi} - i \, \omega t\right)}}{\sin\left({\theta}\right)} - R\left(r\right) \cos\left({\theta}\right) e^{\left(i \, k {\phi} - i \, \omega t\right)} \frac{\partial}{\partial {\theta}}S\left({\theta}\right) \nonumber\\
-& R\left(r\right) e^{\left(i \, k {\phi} - i \, \omega t\right)} \sin\left({\theta}\right) \frac{\partial^{2}}{(\partial {\theta})^{2}}S\left({\theta}\right).
\end{align}
We can find the radial and angular parts of the Klein-Gordon equation after analyzing its structure. This part is semi-automatic and the user should supply some information. 

We first divide the equation by a convenient factor and extract its operands in a vector. The factor is generally the solution Ansatz multiplied by some angular or radial terms. Then we define the separation constant $\lambda_{aux}$. Using a loop over all operands, we decide which one belongs to the radial part and which one belongs to the angular part of the equation. We take the radial derivative of the components and if the result is zero, the component is angular, if it is not zero, we add it to the radial part. We use collection and simplification commands to see the results in a convenient shape.
\begin{lstlisting}[frame=single,firstnumber=last]
# We can analyze the Klein-Gordon equation
# to see how it is separated into
# radial and angular parts.

# Common factors will be divided
# This part should be given by the user
divideKGby=exp(-I*omega*t)*exp(I*k*ph)*sin(th)*R*S
finalKG=expand(KG/divideKGby)

# Extract the operands in the expression:
fkgops=finalKG.operands()

# Find radial and angular parts:
# lambda_aux is the separation constant
var('lambda_aux')
KGradialpart=lambda_aux
KGangularpart=-lambda_aux
for term in fkgops:
	if diff(term,r)==0: 
		KGangularpart=KGangularpart+term
	else: 
		KGradialpart=KGradialpart+term

KGradialpart=expand(KGradialpart*R).simplify_full().collect(R).collect(diff(R,r)).collect(diff(R,r,r))
KGangularpart=expand(KGangularpart*S).simplify_full().collect(S).collect(diff(S,th)).collect(diff(S,th,th))

show('The radial part (variable name is KGradialpart):')
show(KGradialpart)
show('The angular part (variable name is KGangularpart):')
show(KGangularpart)
\end{lstlisting}
The radial part is then found as
\begin{align}
\texttt{KGradialpart}=&\frac{{\left(\omega^{2} r^{3} + 2 \, M \lambda_{\mathit{aux}} - \lambda_{\mathit{aux}} r\right)} R\left(r\right)}{2 \, M - r} + \frac{2 \, {\left(2 \, M^{2} - 3 \, M r + r^{2}\right)} \frac{\partial}{\partial r}R\left(r\right)}{2 \, M - r} \nonumber\\
&+ \frac{{\left(4 \, M^{2} r - 4 \, M r^{2} + r^{3}\right)} \frac{\partial^{2}}{(\partial r)^{2}}R\left(r\right)}{2 \, M - r}, \label{radialkg}
\end{align}
and the angular part is
\begin{align}
\texttt{KGangularpart}=-\frac{\cos\left({\theta}\right) \frac{\partial}{\partial {\theta}}S\left({\theta}\right)}{\sin\left({\theta}\right)} - \frac{{\left(\lambda_{\mathit{aux}} \sin\left({\theta}\right)^{2} - k^{2}\right)} S\left({\theta}\right)}{\sin\left({\theta}\right)^{2}} - \frac{\partial^{2}}{(\partial {\theta})^{2}}S\left({\theta}\right).
\end{align}
We will concentrate on the radial part which is the first step in most quantum gravity problems involving black holes \cite{Birkandan:2011fr}. Further simplifications are obvious but SageMath's \texttt{desolve} command can not give a symbolic solution for the radial equation. However, numerical solutions can be studied using related methods. We will use \texttt{desolve\_system\_rk4} as an example and solve the radial equation numerically. This method uses a fourth--order Runge--Kutta scheme and in fact, the command \texttt{desolve\_system\_rk4} is used as a wrapper for the Maxima command \texttt{rk} \cite{sage}.

Numerical solvers can generally deal with first order equations. Our second order equation will yield two first order differential equations which will be solved simultaneously. We will define the first derivative of the radial function as an auxiliary function
\begin{equation}
\frac{dR}{dr} = R_{aux},
\end{equation} 
and place it in the equation \ref{radialkg} to have
\begin{align} 
\frac{dR_{aux}}{dr} =-\frac{4 \, M^{2} R_{{\rm aux}} + 2 \, r^{2} R_{{\rm aux}} + 2 \, {\left(\lambda_{\mathit{aux}} R - 3 \, r R_{{\rm aux}}\right)} M + {\left(\omega^{2} r^{3} - \lambda_{\mathit{aux}} r\right)} R}{4 \, M^{2} r - 4 \, M r^{2} + r^{3}}.
\end{align}
The following code finds this set of equations as \texttt{radial1} and \texttt{radial2}. The solver can deal with the right hand sides of $\frac{df(x)}{dx}=g(f(x),x)$ type equations. Thus we isolate the derivatives by solving the differential equations as algebraic equations. {\tt radial1} and {\tt radial2} are the right hand sides of our equations to be solved.
\begin{lstlisting}[frame=single,firstnumber=last]
# The auxiliary function
R_aux=function('R_aux')(r)

# Two auxiliary equations
radial1aux=diff(R,r)-R_aux
radial2aux=KGradialpart.subs(diff(R,r)==R_aux).subs(diff(R,r,r)==diff(R_aux,r))

# Right hand sides of the derivatives
radial1=(solve(radial1aux==0,diff(R,r)))[0].right()
radial2=(solve(radial2aux==0,diff(R_aux,r)))[0].right()

# Getting the outputs in input format
# (Outputs will be copied)
print(radial1)
print(radial2)
\end{lstlisting}
In their present forms, $R$ and $R_{aux}$ are defined as functions. However, in order to use \texttt{desolve\_system\_rk4} we need to define unknowns as variables. To do this, we need to copy the outputs for \texttt{radial1} and \texttt{radial2} and change them accordingly. We display the outputs by {\tt print} command to see them in the input form which enables us to copy them easily.

After a formal analysis of the radial part, one can see that the radial solution can be given in terms of confluent Heun functions \cite{Vieira:2016ubt}. General and confluent forms of the Heun function are encountered in many applications in physics, especially as solutions of the wave equations \cite{ronveaux,slav,hortacsuheun,Birkandan:2006ac,Birkandan:2016dsr,Fiziev:2011mm}.

Numerical computation of the general and confluent Heun functions are adapted by Oleg V. Motygin for GNU Octave \cite{oleg1,oleg2}. However, no freely available packages or modules can give closed symbolic solutions of these equations. Therefore we cannot compare our result with the full analytical solution. The asympotic form of the confluent Heun function is given in reference \cite{Vieira:2016ubt} as
\begin{equation}
R_{\ell}=\frac{C_{\ell}}{r}\sin\bigg[\omega r+2M\omega \ln(r) - \frac{\ell \pi}{2} +\arg \big(\Gamma(\ell+1-2iM\omega)\big) \bigg],
\end{equation}
where $\lambda_{aux}=\ell(\ell+1)$.

We first import the differential equation solver from {\tt sage.calculus.desolvers}. We then define the variables (including the unknown functions) and the equations and copy the equations from the outputs of the code above. We give some numerical values to the parameters arbitrarily and solve the system for some arbitrary set of initial conditions. In our example we have $R(r=0.3)=1$ and $\frac{dR}{dr}=R_{aux}(r=0.3)=0.5$. The solution is stored in {\tt radsol}.

{\tt radsol} has the structure $[r, R, R_{aux}]$. {\tt points=[[i,j] for i,j,k in radsol]} command creates {\tt [i,j]} (or $[r, R]$) points for plotting. {\tt radialsolution} stores the plot of the numerical result.

\begin{figure}[H]
	\centering
	\includegraphics[scale=0.6]{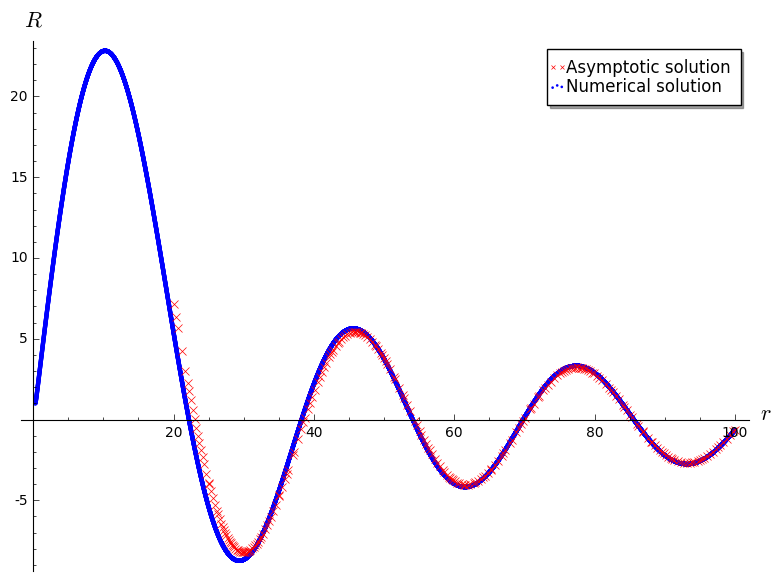} \caption{The radial part of the Klein--Gordon solution.}%
	\label{fig:fig1}%
\end{figure}

In the next part, we define the asymptotic form of the analytical solution and plot it for the same parameter set (we take $C_{\ell}=250$ to match the amplitude). We display both plots together to show the agreement.
\begin{lstlisting}[frame=single,firstnumber=last]
# Import the solver
from sage.calculus.desolvers import desolve_system_rk4

# Define unknowns as variables
var('R,R_aux,r')

# Define equations
radial1=R_aux
radial2=-(4*M^2*R_aux + 2*r^2*R_aux + 2*(lambda_aux*R - 3*r*R_aux)*M + (omega^2*r^3 - lambda_aux*r)*R)/(4*M^2*r - 4*M*r^2 + r^3)

# Substitute numerical values for parameters
radial2=radial2.subs(M=0.1,omega=0.2,k=2.0,lambda_aux=2.0)

# Solve the system and plot the solution
radsol=desolve_system_rk4([radial1,radial2],[R,R_aux],ics=[0.3,1,0.5],ivar=r,end_points=100,step=0.01)
points=[[i,j] for i,j,k in radsol]
radialsolution=list_plot(points,axes_labels=['$r$','$R$'],legend_label='Numerical solution')

# Asymptotic form of the analytic solution
var('Cl,L')
Rasym=Cl*(1/r)*sin(omega*r+2*M*omega*log(r)-(L*pi/2)+arg(gamma(L+1-2*I*M*omega)))
Rasymnum=Rasym.subs(Cl=250,M=0.1,omega=0.2,L=1.0)
asympplot=plot(Rasymnum,(r, 20, 100),linestyle='',marker='x',color='red',legend_label='Asymptotic solution')

# Display both plots
show(asympplot+radialsolution) (*@\label{ex4}@*)
\end{lstlisting}
The plot \ref{fig:fig1} of the solution gives an idea on the behavior of the radial function. Numerical solution and the asymptotic form of the analytical solution are plotted together to show that they agree for large $r$.
%%%%%%%%%%%%%%%%%%%%%%
\subsection{Visualizing geodesics}
The geodesic motion in a spacetime can be described by the Hamilton--Jacobi equation \cite{chandr}.
\begin{equation}
\frac{\partial S}{\partial \eta}-\frac{1}{2}g^{\mu \nu}\frac{\partial S}{\partial x^{\mu}}\frac{\partial S}{\partial x^{\nu}}=0,
\end{equation}
where $S$ denotes the Hamilton's principal function, $g^{\mu \nu}$ is the inverse metric and $\eta$ is an affine parameter. The orbital equations then found using
\begin{equation}
\frac{d x^{\mu}}{d \eta}+g^{\mu \nu}\frac{\partial S}{\partial x^{\nu}}=0.
\end{equation}
This formalism gives first order differential equations which are needed in numerical schemes as mentioned in the previous section.

The Hamilton's principal function is decomposed as
\begin{equation}
S(\eta,t,r,\theta,\phi)=\frac{m^2}{2}\eta-Et+L\phi+F(r)+G(\theta),
\end{equation}
for the Schwarzschild spacetime. For the equatorial geodesics we take $\theta=\pi/2$ and thus $G(\theta)=0$. The metric components also change according to this constraint.

Before beginning our analysis, it should be emphasized that decomposition of the Hamilton's principle function $S$ (and the scalar field $\Phi$ in the previous section) is not evident and needs a detailed analysis in general spacetimes, which is beyond the scope of our work. Moreover, analysis of geodesics is a very detailed study and our primitive example here aims only to visualize some geodesics.

The user should execute the lines (\ref{ex1}-\ref{ex2}) for the definition of the spacetime before running the codes below. 

We start by defining our variables and functions. We also calculate the inverse metric as needed in the equations. After giving the Hamilton's principle function Ansatz, we calculate the Hamilton--Jacobi equation in two loops and set it to the variable {\tt HJfull}. We call it {\tt HJfull} as we have not put any conditions ($\theta=\pi/2$, etc.) on the equation yet.

The right hand sides (as needed by differential equation solvers) of the geodesic (orbital) equations are then calculated (using nested loops) and put in the vector {\tt geodeqnrhs}. We display the equations both in the {\LaTeX} format and as a vector.
\begin{lstlisting}[frame=single][firstnumber=metricdef]
# Define variables, functions and calculate inverse metric (*@\label{ex5}@*)
var('eta,m,E,L,S,HJfull')
F=function('F')(r)
G=function('G')(th)
ginv = g.inverse()

# Define the principal function Ansatz
S=((eta*m^2)/2)-E*t+L*ph+F+G

# Calculate the Hamilton-Jacobi equation
HJfull=0
for i in range(len(BL[:])):
	for j in range(len(BL[:])):
		HJfull=HJfull+ginv[i,j].expr()*diff(S,BL[i])*diff(S,BL[j])
HJfull=(diff(S,eta)-(1/2)*HJfull)
show('The Full Hamilton-Jacobi equation (variable name is HJfull):')
show(HJfull)

show('The geodesic equations in LaTeX form (variable name is geodeqnrhs):')
geodeqnrhs=zero_vector(SR, len(BL[:]))

for mu in range(len(BL[:])):
	for nu in range(len(BL[:])):
		geodeqnrhs[mu]=geodeqnrhs[mu]-(ginv[mu,nu].expr())*diff(S,BL[nu])
	writeresult='D[0](%s)($\eta$) = $%s$' %(BL[mu],latex(geodeqnrhs[mu]))
	show(writeresult)

show('Right hand sides of the geodesic equations as a vector')
show(geodeqnrhs)
\end{lstlisting}
The code above is general and it can work for any spacetime if the Hamilton's principal function is given accordingly. The codes below depend on the Schwarzschild metric.

Conventionally, we take $\theta=\pi/2$ to find the equatorial geodesics. We call the Hamilton--Jacobi equation with this constraint as {\tt HJst} and the right hand sides of the orbital equations as {\tt geodeqnrhsst}. We substitute $\theta=\pi/2$ and $G(\theta)=0$ in the equations found in the general code.
\begin{lstlisting}[frame=single][firstnumber=last]
# Take theta = pi/2:
var('HJst')

# Hamilton-Jacobi equation for theta=pi/2 (variable name is HJst)
HJst=(HJfull.subs(diff(G)==0)).subs(th=pi/2)
show(HJst)

# Right hand sides of the geodesic equations for theta=pi/2 
# (variable name is geodeqnrhsst)
geodeqnrhsst=(geodeqnrhs.subs(diff(G)==0)).subs(th=pi/2)
show(geodeqnrhsst)
\end{lstlisting}
We find
\begin{equation}
{\tt HJst}=\frac{m^2}{2}+\frac{E^2r}{2(2M-r)}+\frac{L^2}{2r^2}-\frac{2M-r}{2r}    \bigg(\frac{dF(r)}{dr} \bigg)^2,
\end{equation}
and the nonzero components of {\tt geodeqnrhsst} are
\begin{align}
\frac{dt}{d\eta}&=-\frac{Er}{2M-r}, \\
\frac{dr}{d\eta}&=-\frac{2M-r}{r}\bigg(\frac{dF(r)}{dr} \bigg), \\
\frac{d\phi}{d\eta}&=\frac{L}{r^2}, 
\end{align}
We are now ready to solve the orbital equations. Here, we will use another solver, {\tt desolve\_odeint} by importing from {\tt sage.calculus.desolvers}. This solver uses {\tt scipy.integrate} module of Python. We will plot the geodesic curve, and a black disc with a radius equal to the event horizon radius in order to indicate the black hole. Thus, we import {\tt Circle} from {\tt sage.plot.circle}. We define the variables and give them arbitrary numerical values. We isolate $\frac{dF(r)}{dr}$ as {\tt derofradfun} from {\tt HJst} and place it in $\frac{dr}{d\eta}$ equation. 

We would like to solve time-dependent $\frac{dr}{dt}$ and $\frac{d\phi}{dt}$ equations, instead of the equations with affine-parameter ($\eta$) dependence. Thus, we divide $\frac{dr}{d\eta}$ and $\frac{d\phi}{d\eta}$ by $\frac{dt}{d\eta}$, and set them as our equations: {\tt geodeqn1} and {\tt geodeqn2}. 

\begin{figure}[H]
	\centering
	\includegraphics[scale=0.4]{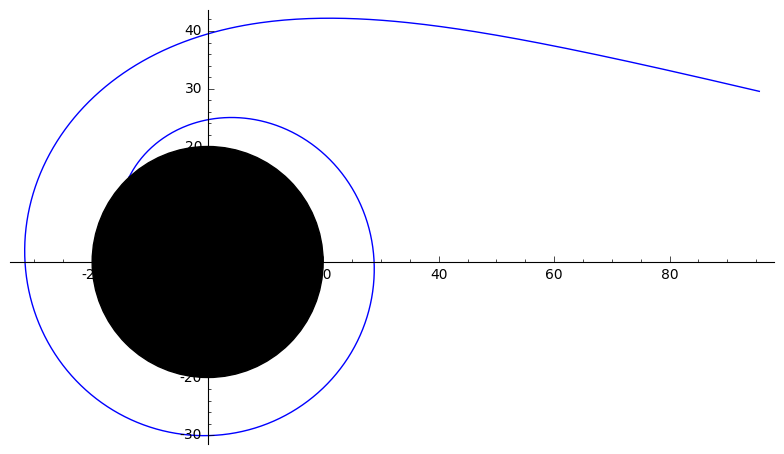} \caption{Example of a null geodesic for the Schwarzschild black hole ($\theta=\pi/2$).}%
	\label{fig:fig2}%
\end{figure}

\begin{figure}[H]
	\centering
	\includegraphics[scale=0.4]{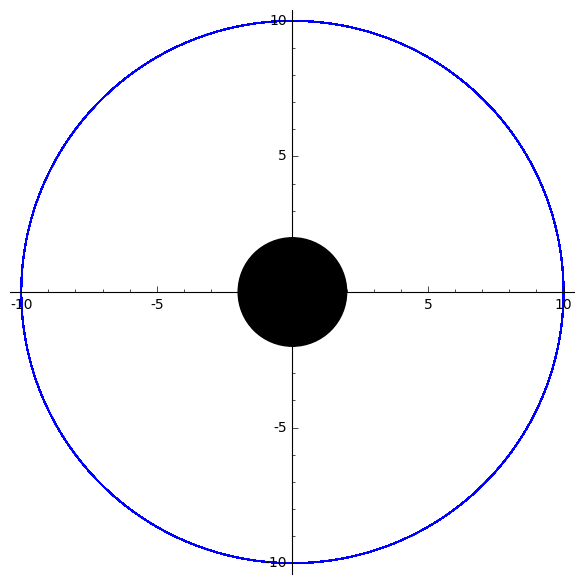} \caption{Stable circular orbit for the Schwarzschild black hole for a timelike particle ($\theta=\pi/2$).}
	\label{fig:fig3}
	%m_aux=1;M_aux=1;L_aux=4;E_aux=1.0;step_size=0.1;
	%eta_end=10000;r_initial=5*M_aux;ph_initial=0.3;
\end{figure}

\begin{figure}[H]
	\centering
	\includegraphics[scale=0.4]{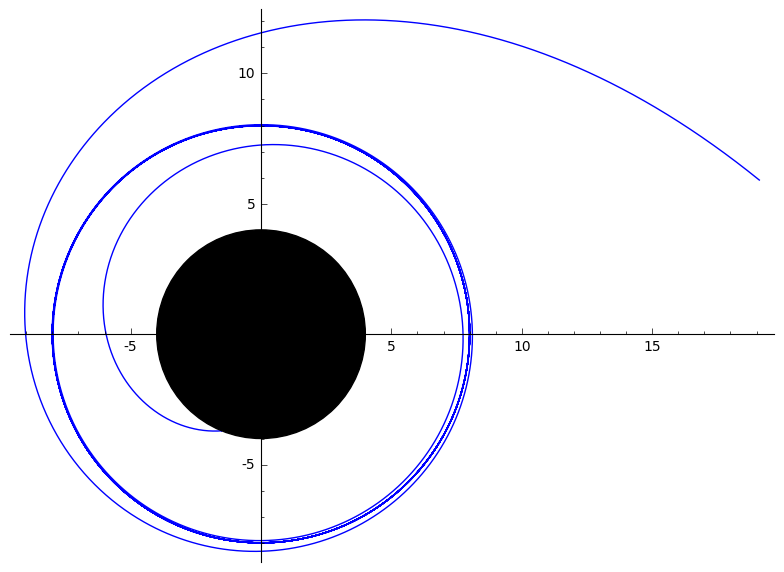} \caption{Example of a timelike geodesic for the Schwarzschild black hole ($\theta=\pi/2$).}%
	\label{fig:fig4}%
	%m_aux=1;M_aux=2;L_aux=8.0;E_aux=1.0;step_size=0.1;
	%eta_end=10000;	r_initial=10*M_aux;	ph_initial=0.3;
\end{figure}

We solve our equations for arbitrarily set initial conditions and put the results in the variable {\tt sol}. The $0^{th}$ column of {\tt sol} carries the $r$ values and the $1^{st}$ column carries the $\phi$ values. Using these, we generate $\{x,y\}$ points, where $x=rcos(\phi)$ and $y=rsin(\phi)$. The {\tt line} plot of these points forms the geodesic curve. We plot this curve and the black disc (a circle with parameters {\tt fill=True,rgbcolor=`black'}) together to display the behavior.
\begin{lstlisting}[frame=single][firstnumber=last]
# Importing the solver
from sage.calculus.desolvers import desolve_odeint
#Importing circle for visualizing the black hole
from sage.plot.circle import Circle

var('m_aux,L_aux,E_aux,M_aux,r_initial,ph_initial,eta_end,step_size')

##########################
# Variables
m_aux=1 # Either 1 (timelike) or 0 (null).
M_aux=1
L_aux=4
E_aux=1.0
step_size=0.1
eta_end=500
r_initial=2.1*M_aux
ph_initial=0.3
##########################

# Derivative of F (the radial function)
# (variable name is derofradfun)
# We will use the first root
derofradfun=solve(HJst,diff(F,r))

# Define equations to solve 
# dr/dt = geodeqn1
# dphi/dt = geodeqn2
geodeqn1=((geodeqnrhsst[1]/geodeqnrhsst[0]).subs(diff(F,r)==derofradfun[1].rhs())).subs(E=E_aux,L=L_aux,m=m_aux,M=M_aux)
geodeqn2=(geodeqnrhsst[3]/geodeqnrhsst[0]).subs(E=E_aux,L=L_aux,m=m_aux,M=M_aux)

# Solve the equations
sol=desolve_odeint([geodeqn1,geodeqn2],[r_initial,ph_initial],srange(0,eta_end,step_size),[r,ph])
p=line(zip(sol[:,0]*cos(sol[:,1]),sol[:,0]*sin(sol[:,1])))

# Plot the black hole as a circle 
# Show the geodesics and the circle on the same plot
C=circle((0,0),2*M_aux,fill=True,rgbcolor='black')
show(C+p) (*@\label{ex6}@*)
\end{lstlisting}
The figure \ref{fig:fig2} shows an example of a null geodesic ($m=m_{aux}=0$) for an arbitrary set of parameters. Figure \ref{fig:fig3} shows the stable circular orbit for a timelike particle ($m=m_{aux}=1$) and Figure \ref{fig:fig4} shows another example of a timelike geodesic.

%%%%%%%%%%%%%%%%%%%%%%
\section{Conclusion}
We studied three computer algebra systems which allow tensor manipulation for GR calculations, and showed how one can define a spacetime, calculate Christoffel symbols, Ricci tensor and Einstein tensor for this spacetime using the specialized commands of these systems. We used SageMath with its tensor manipulation and differential geometry module SageManifolds, Maxima with its tensor component manipulation package ctensor and the Python language with the GraviPy module which runs on the symbolic analysis module SymPy. A benchmark for these systems is also provided.

In the next part, our main focus was the SageMath+SageManifolds system to perform two examples: Massless Klein--Gordon equation calculations and geodesic visualization for the Schwarzschild geometry. Primarily, the general codes that can run for general spacetimes are given, and then the analysis is specialized for the Schwarzschild case.

In our first example, we derived the massless Klein--Gordon equation as a partial differential equation using our code. We then separated it to the radial and angular parts and analyzed the radial part in detail. We confirmed our numerical result graphically using the asymptotic form of the analytical result. This part involves a direct numerical integration of a confluent Heun equation resulting from the radial part of the Klein--Gordon equation and its comparison with the asymptotic analytical result which is rare in the literature.

In the second example, the orbit of a null or timelike particle around a Schwarzschild black hole was constructed from the geodesic equations. There are multiple methods of doing that, and among them, the chosen one here is the Hamilton--Jacobi formalism as it yields first order differential equations that can be treated easier than the second order equations on the computer. The orbits are visualized using the plotting tools of SageMath. Our example is far from a complete analysis of geodesics although it provides a basis for such a study.

A computation system should be chosen according to the needs of the problem. For example, if a research topic depends on symbolic manipulation of special functions of mathematical physics, commercial packages are inevitable for most of the cases in the present situation. However, many problems do not need such specialized calculations and numerical analysis is sufficient to see the result. As seen in the examples, in many cases freely available packages are capable of forming a complete system for scientific problems.
%%%%%%%%%%%%%%%%%%%%%%%%%%%%
\section*{Acknowledgement}
We would like to thank Profs. Ne\c{s}e \"{O}zdemir, Durmu\c{s} Ali Demir and
\'{E}ric Gourgoulhon for stimulating discussions. We also thank our anonymous referee for the constructive comments which helped us to improve the manuscript. This work is partially supported by T\"{U}B\.{I}TAK, the Scientific and Technological Council of Turkey.
\end{document}